%% file: main.tex
\documentclass{egpubl}
\usepackage{eurovis2022}

\SpecialIssuePaper         

\usepackage[T1]{fontenc}
\usepackage{dfadobe}  

\usepackage{cite}  
\BibtexOrBiblatex
\electronicVersion
\PrintedOrElectronic
\ifpdf \usepackage[pdftex]{graphicx} \pdfcompresslevel=9
\else \usepackage[dvips]{graphicx} \fi

\usepackage{egweblnk}
\graphicspath{{figures/}{pictures/}{images/}{./}} 

\usepackage{microtype}                 
\PassOptionsToPackage{warn}{textcomp}  
\usepackage{textcomp}                  
\usepackage{mathptmx}                  
\usepackage{times}                     
\usepackage{cite}                      
\usepackage{tabu}                      
\usepackage{booktabs}                  
\usepackage{xcolor}
\usepackage{url}
\usepackage{amsfonts}
\usepackage{amsmath}
\usepackage{amssymb}
\usepackage{enumitem}
\usepackage{flushend}

\usepackage{color}
\definecolor{RED}{rgb}{1, 0, 0}
\definecolor{GREEN}{rgb}{0, 0.6, 0}
\definecolor{BLUE}{rgb}{0, 0.4, 1}
\definecolor{GRAY}{rgb}{0.5, 0.5, 0.5}

\definecolor{RED}{rgb}{1, 0, 0}
\definecolor{GREEN}{rgb}{0, 0.6, 0}
\definecolor{BLUE}{rgb}{0, 0.4, 1}
\definecolor{GRAY}{rgb}{0.5, 0.5, 0.5}

\title[Characterizing Grounded Theory Approaches in Visualization]%
      {Characterizing Grounded Theory Approaches in Visualization}

\author[A. Diehl et al.]
{\parbox{\textwidth}{\centering%
        A. Diehl$^1$, 
        A. Abdul-Rahman$^2$\orcid{0000-0002-6257-876X},
        B. Bach$^3$,
        M. El-Assady$^4$,
        M. Kraus$^5$,
        R. S. Laramee$^6$,
        D. A. Keim$^5$,
        M. Chen$^7$\orcid{0000-0001-5320-5729}
        }
        \\
{\parbox{\textwidth}{\centering%
        $^1$University of Zurich, Switzerland \quad
        $^2$King's College London, UK \quad
        $^3$Edinburgh University, UK\\
        $^4$ETH AI Center, Zurich, Switzerland \quad
        $^5$University of Konstanz, Germany \quad
        $^6$Nottingham University, UK \quad
        $^7$University of Oxford, UK
        }
}
}

\begin{document}

\teaser{
    \centering
    \includegraphics[width=176mm]{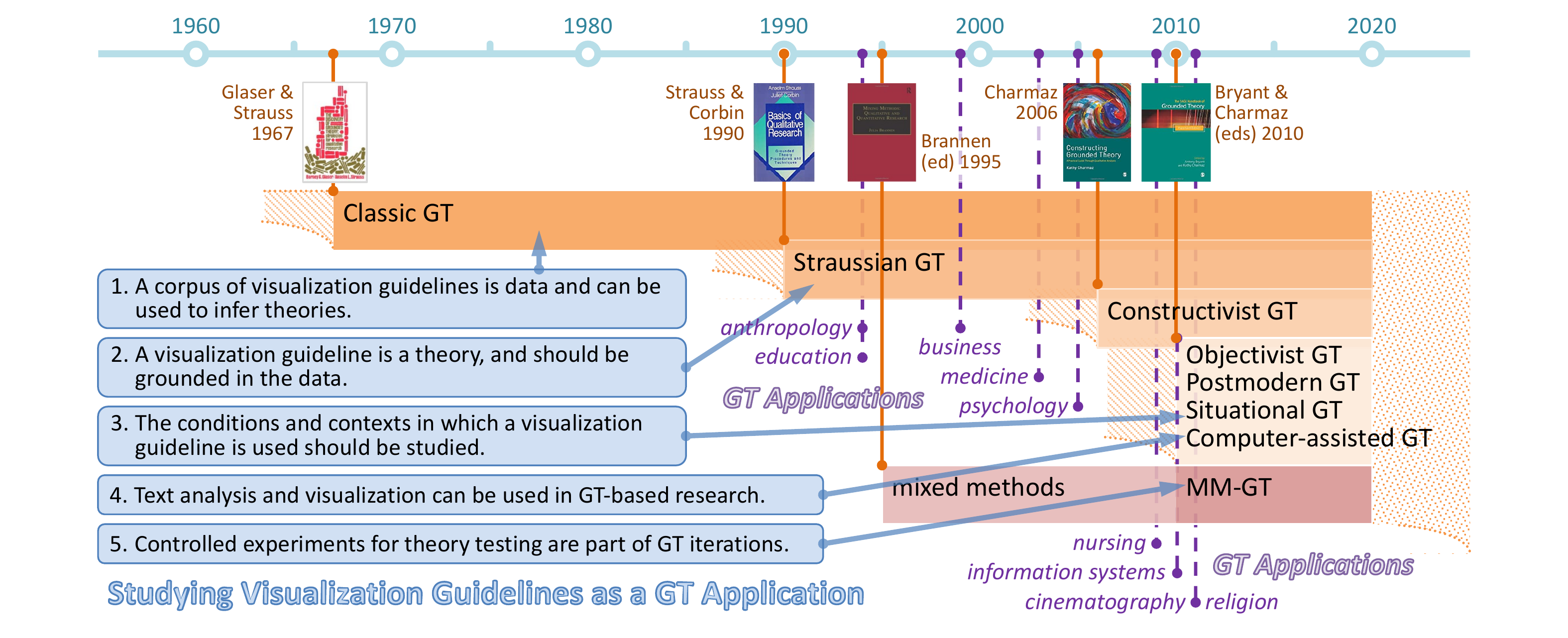}
    \caption{Over the last five decades, the scope of GT has been broadened gradually. Many methods used in VIS for studying discourse corpora fall within the scope of GT, and the combined uses of these methods can strengthen the GT applications in VIS.}
    \label{fig:GTandVis}
}

\maketitle
\begin{abstract}
Grounded theory (GT) is a research methodology that entails a systematic workflow for theory generation grounded on emergent data. 
In this paper, we juxtapose GT workflows with typical workflows in visualization and visual analytics, shortly VIS, revealing the characteristics shared by these workflows. We explore the research landscape of VIS to observe where GT has been applied to generate VIS theories, explicitly as well as implicitly.
We discuss ``why'' GT can potentially play a significant role in VIS.
We outline a ``how'' methodology for conducting GT research in VIS, which addresses the need for theoretical advancement in VIS while benefiting from other methods and techniques in VIS. We exemplify this ``how'' methodology by adopting GT approaches in studying the messages posted on VisGuides --- an Open Discourse Forum for discussing visualization guidelines.

\ccsdesc[500]{Human-centered computing~Visualization theory, concepts and paradigms}

 \printccsdesc   
\end{abstract} 

\input{sections/1NewIntroduction}

\input{sections/2NewRelatedWork}
\input{sections/3NewGT}

\input{sections/4NewWorkflows}
\input{sections/5NewWhere}
\input{sections/6NewWhyHow}
\input{sections/7NewConclusions}


\bibliographystyle{eg-alpha-doi}
\bibliography{visguides}







\end{document}

%% file: sections/1NewIntroduction.tex
\section{Introduction}
\label{sec:NewIntro}

\emph{Grounded theory} (GT) is a well-established methodology in social science to scrutinize a postulated theory by \textit{grounding} it in data.
As depicted in Figure \ref{fig:GTandVis}, GT was first proposed by Glaser and Strauss in 1967 for sociological research \cite{Glaser:1967:book}, and its scope has been further broadened since 1990. In addition to disciplines in social science, it has been applied to medicine, nursing, and computer science.

At a high level, GT workflows bear a strong resemblance to visualization and visual analytics (VIS) workflows. Both types of workflow aim to derive new knowledge from data (e.g., discovering theories in GT), and enable human users to make data-informed decisions (e.g., determining categories for a concept).
At a low level, most GT workflows are yet to benefit significantly from VIS techniques and tools.
This motivates us with the first related question: \emph{Should and can VIS techniques be deployed in GT workflows?}

Meanwhile, while GT methods have been applied to study VIS phenomena, the number of VIS publications that explicitly attributed the work to GT is relatively small. This motivates us with the second related question: \emph{Should and can GT methods be more widely deployed in studying VIS phenomena?}


This discourse paper, which follows the tradition of many discourse papers in VIS (Section~\ref{sec:NewRelatedWork}), reports a qualitative analytical investigation into the above two questions.
In our investigation, we first research the GT literature and map out different schools of thought in GT. We conclude that classic GT would have welcomed VIS if such technology had been available in the 1960s when GT was first proposed. We find that modern GT encompasses computer-assisted GT methods as well as empirical studies (Section~\ref{sec:NewGT}).
We then compare a typical GT workflow with a typical VA workflow and a typical hypothesis-testing workflow. We examined the similarity and difference among them, and outlined a VIS4GT workflow that falls within the scope of modern GT and is consistent with the principles of classic GT (Section~\ref{sec:NewWorkflows}).
We thus concluded that VIS techniques \textbf{should} be deployed in GT workflows.

Building on the above discourse, we considered several areas where VIS phenomena may be studied using GT methods, and discovered that although the number of GT papers in VIS is small, there have been many VIS research papers exhibiting GT-like research methods. We thus concluded that GT methods \textbf{should} and \textbf{can} be widely deployed in studying VIS phenomena (Section~\ref{sec:NewWhere}).
Following a brief discourse on the benefit of using GT in VIS, we analysed our own experience of using GT in studying the discourse data on visualization guidelines, and enriching our categorization effort with VIS techniques and empirical studies (Section \ref{sec:NewWhyHow}). We concluded that VIS techniques \textbf{can} be deployed in GT workflows.

%% file: sections/2NewRelatedWork.tex
\section{Related Work}
\label{sec:NewRelatedWork}
%
In this section, we provide a summary review of discussions and discourse papers in VIS. We then briefly review VIS publications where GT methods were mentioned. 
\subsection{Discourse-based Research in VIS}
In VIS, a good number of discourse papers have had significant influence on the subject development of VIS, including, e.g., Van Wijk's discourse on the value of visualization \cite{van2005value} and Munzner on design workflows \cite{munzner2009nested}.
The discourse on visual analytics (VA) by Thomas and Cook \cite{Thomas:2005:book} and Keim et al. \cite{Keim:2008:bookCh,keim2010mastering} shaped the subject of visual analytics today.
In general, discourse-based research enables researchers to translate their experience and observations into insight on a specific topic, engage in scholarly discussions and critiques, and propose new methodologies and research directions. While it is not feasible to include many discourse papers in this brief review, here we highlight a number of important papers representing different research fora.

\textbf{Discussions and Critiques.}
Tory and M\"{o}ller discussed the subject division in VIS \cite{tory2004rethinking} and argued for the importance of the human as a factor to consider during the visualization design process\cite{tory2014user}.
Chen et al. questioned if the term \emph{insight} answers the question ``what is visualization really for?'' \cite{chen2014visualization}.
Streeb critiqued some existing arguments about the biases in VIS \cite{streeb2018biases}.
Chen et al. provided analytical reasoning about the cost-benefit of deploying VIS applications in VR environments \cite{chen2018information}.
Dimara and Perin offered their answers to the question ``what is interaction for VIS?'' \cite{dimara2019interaction}.
There are also discourse papers on discourse. For example, Streeb et al. examined the numerous arguments about the question ``why we need visualization'' or ``visual analytics'' \cite{streeb2019visualize,streeb2021visualize}.
Chen and Edwards examined a collection of schools of thought or ``isms'' in VIS \cite{chen2020isms}. 

\textbf{Conceptual Models.} Built on the VIS pipeline proposed by Van Wijk \cite{van2005value} and VA pipeline by Keim et al. \cite{Keim:2008:bookCh}, 
Sacha et al. proposed an extended VA pipeline to include analytical reasoning, knowledge acquisition and decision making processes in the mind.

\textbf{Methodologies.} Following Munzner's discourse on design workflows, there have been a series of discourse papers on
design study methodologies \cite{sedlmair2012design,bigelow2014reflections},
using guidelines in design \cite{Meyer:2015:IV},
criteria for rigorous design studies \cite{meyer2019criteria},
problem-driven visualization \cite{marai2017activity,hall2019design},
discourse in design processes \cite{beecham2020design}, and
best design practices \cite{parsons2020design,crisan2020passing,parsons2021understanding}.
Lam et al., Isrnberg et al., Elmqvist, Saket et al. provided a series of discourse on evaluation methodologies in VIS \cite{isenberg2013systematic,lam2011empirical,saket2016beyond,elmqvist2015patterns}.
Brehmer et al., Rind et al., and Kerracher et al. proposed methodologies for tasks abstraction in VIS \cite{brehmer2013multi,kerracher2017constructing,rind2016task}.

\textbf{Research Agenda.} Johnson's discourse on scientific visualization research agenda outlined a number of important unsolved problems \cite{Johnson:2004:CGA}.
Chen et al. mapped out the pathways for theoretical advances in VIS \cite{chen2017pathways}.
Bradley et al. proposed interdisciplinary research agenda in conjunction with digital humanity (DH), including GT methods in VIS \cite{Bradley:2018:CGA}.

These discourse papers confirm the tradition and the need for discourse-based research in VIS. We can infer the need for using GT in studying discourse in VIS.
This work provides a new discourse on an important interdisciplinary topic between VIS and social science. In particular, we discuss the need for increasing the uses of GT methods in VIS, and propose a VIS-assisted GT workflow that can benefit from VIS techniques and tools as well as empirical studies.

\subsection{Applications of GT in VIS}
GT has been accepted as a major qualitative research methodology by the field of human-computer interaction for at least two decades \cite{Creswell:2002:book}.
Knigge and Cope \cite{knigge2006grounded} proposed an integrated analytical method that combines GT, geographic information systems, and ethnography in the area of environment and planning, and coined the term \emph{Grounded Visualization}. 
In the context of visualization, Kandogan and Lee applied GT methods to study visualization guidelines \cite{kandogan2016grounded}.
Isenberg et al. used GT for evaluation in VIS and referred to it as \emph{Grounded Evaluation} \cite{isenberg2008grounded}.
Seldmair et al. proposed a new design study methodology and compared it with other methodologies such as ethnography, action research, and GT \cite{sedlmair2012design}.
Lee et al. used GT to construct a grounded model of guidelines for making sense of unfamiliar visualizations~\cite{lee2015people}.
Bigelow et al. used GT to investigate how VIS designers work with data \cite{bigelow2020guidelines}. 
Lundgard and Satyanarayan used GT to understand the semantics of natural language descriptions and derived a four-level model of semantic content \cite{lundgard2021accessible}.
Diehl et al. collected discourse on different visualization guidelines and practices and applied GT to study the collected corpus \cite{Diehl:2020:arXiv, Diehl:2018:EuroVis}.

In Section \ref{sec:NewWhere}, we will discuss previous work in the VIS literature even further, which may feature GT-link workflows. In Section \ref{sec:NewWhyHow}, we will use the work by Diehl et al. \cite{Diehl:2020:arXiv} to inform the discourse on how VIS techniques and VIS empirical studies may be used in GT workflows.

%% file: sections/3NewGT.tex
\section{Grounded Theory: Terminology and Concepts}
\label{sec:NewGT}

Grounded Theory (GT) is a research methodology centered on the principle that theories must be grounded in the data%
\footnote{GT itself is not a theory. In this paper, the term ``a theory'' means a theoretical postulation that is to be formulated and evaluated using GT methods, unless it is stated explicitly as a confirmed theory.}.
GT was first proposed by Glaser and Strauss for sociological research \cite{Glaser:1967:book} with the goal of constructing and scrutinizing theories from data.
Applications of GT normally feature two major processes: (i) data collection, e.g., in the form of examples, counter examples, and case studies; and (ii) data analysis where data is examined through methods such as as \emph{categorization}, \emph{coding}, \emph{constant comparative analysis}, \emph{negative case analysis}, and \emph{memoing}.
The ultimate goal of GT is to derive a theory from raw data and/or scrutinize a proposed theory systematically by constantly and continuously sampling the data space, analyzing the data captured, and refining the theory until it reaches \emph{theoretical saturation}.
The first two columns of Table~\ref{tab:GTmethods} summarize and compare these methods.

In addition, in the original proposal of the classic GT \cite{Glaser:1967:book}, Glaser and Strass articulated the need for computing statistical measures for the categorized data, and for placing the statistical measures in tables intelligently such that the relationships among different categories can be observed.
These two methods remind us of the basic activities in ``analytics'' and ``visualization'' in VIS, while we can easily understand why these two words were absent in the original proposal of GT \cite{Glaser:1967:book} due to the historical context.
We therefore added two extra rows at the end of Table \ref{tab:GTmethods} to highlight the presence of ``analytics'' and ``visualization'' in GT.  
The formal descriptions of these methods and principles, which was compiled based on the work of Willig~\cite{Willig:2013:book}, can be found in Appendix A.

\begin{table*}[t!]
  \centering
  \caption{A summary of the principles and methods of grounded theory, and an outlook on how computer-assisted implementation may complement the traditional implementation of the grounded theory (GT) methodology \cite{Willig:2013:book}. In the last two rows, we use place ``analytics'' and ``visualization'' in quotation marks because they are not the original terms used in traditional implementation of GT methods.}
  \label{tab:GTmethods}
  
  \begin{tabular}{@{}p{3.2cm}@{\hspace{2mm}}p{6cm}@{\hspace{3mm}}p{8cm}@{}}
  \textbf{Methods and Principles} & \textbf{Traditional Implementation} &  \textbf{Computer-assisted Implementation}\\
  \toprule
  \textbf{Categorization}
  & Close reading;
    category identification.
  & Distant reading with statistics and statistical graphics;
    cluster and similarity analysis;
    text analysis \& visualization.\\
  \midrule
  \textbf{Theorization}
  & Taxonomy \& ontology construction;
    entity relationships (e.g., correlation, association, causality, etc.);
    guidelines;
    conceptual models.
  & Visual analytics;
    machine learning;
    algorithmic taxonomy \& ontology learning;
    correlation, association analysis \& causality analysis;
    dimensionality reduction;
    modelling \& simulation.\\    
  \midrule
  \textbf{Coding}
  & Close reading and labelling;
    open coding;
    axial coding;
    selective coding.
  & Text and discourse analysis;
    text visualization;
    similarity \& association analysis;
    network visualization;
    ontology mapping.\\
  \midrule
  \textbf{Comparative Analysis}
  & Iterative close reading;
    comparing different options of category abstraction;
    continuous refinement of the categorization scheme;
    proposing new categories or categorization schemes. 
  & Iterative and continuous effort for distant reading with statistics and statistical graphics; cluster \& similarity analysis; topic modelling; text visualization; algorithmic taxonomy \& ontology learning.\\
  \midrule
  \textbf{Negative Case Analysis}
  & Close reading;
    negative case identification \& analysis. 
  & Algorithmic outlier analysis \& anomaly detection.\\
  \midrule
  \textbf{Memoing}
  & Memo-writing;
    sketch-drawing;
    systematically recording the ideas and actions related to the development of a theory. 
  & Online discussion forum;
    crowd sourcing;
    provenance visualization;
    data collection from social media.\\
  \midrule
  \textbf{Theoretical Sensitivity}
  & Testing a theory against old and new data (i.e., ``Interact'' with data);
    analysis of applicability, negative cases, and possible new categories. 
  & Computer-assisted data collection;
    data analysis \& visualization.\\
  \midrule
  \textbf{Theoretical Sampling}
  & Data collection;
    empirical studies (e.g., controlled experiments, surveys, focus groups, field observation, etc.);
  interview transcription.
  & Online discussion forum,
    crowd sourcing;
    data collection from social media;
    computer-assisted experiments;
    computer-mediated group activities;
    observation of online activities.\\
  \midrule
  \textbf{Theoretical Saturation}
  & Data collection and analysis until theoretical saturation has been achieved.
  & Computer-assisted platforms for facilitating the longevity and provenance of the GT processes.\\
  \midrule
  \textbf{``Analytics''}
  & Computing statistical measures of categorized data.
  & Statistical inference; data mining; data modelling; visual analytics.\\
  \midrule
  \textbf{``Visualization''}
  & Displaying statistical measures in tables with intelligent alignment.
  & General-purpose and special-purpose visual representations; interactive visualization; visual analytics.\\
  \bottomrule
  \end{tabular}
\vspace{-4mm}
\end{table*}

As illustrated in Figure \ref{fig:GTandVis}, the scope of GT has been broadened over the past five decades \cite{Kenny:2014:QR}.
The original version of GT asserted that a theory must be ``naturally'' emerged from the data, and GT researchers must be open-minded and must abstain from existing theories in the literature.
1990 saw the first major extension of the scope.
Strauss (one of the original proposers) and Corbin challenged this assertion, reinterpreting ``open mind'' as not ``empty mind'' \cite{Strauss:1990:book}.
This broadened the scope of GT beyond the restriction of studying ``naturally-emerged'' theories only.
In 2006, Charmaz, a student of Strauss and Glaser, further broadened the scope of GT by allowing GT researchers to use their prior knowledge more actively, especially in coding, sampling, raising questions, and designing further iterations \cite{Charmaz:2006:book}.
Meanwhile, Glaser argued to preserve the original scope of GT, which is now referred to as \emph{Classic GT}, while the two extended versions are referred to as \emph{Straussian GT} and \emph{Constructivist GT} respectively. 
The insistence for maintaining the ``pure GT'' is often referred to as \emph{Glaserian GT}.

Several other extensions were proposed before or around 2010.
The SAGE Handbook of Grounded Theory edited by Bryant and Charmaz \cite{Bryant:2010:book} consists of 27 articles on GT, featuring different versions and applications of GT.
In their introductory article of the handbook \cite{Bryant:2010:bookCh1}, Bryant and Charmaz explicitly defined ``grounded theory method as a family of methods,'' and marked the handbook as a celebration of difference.
In another article, Denzin identified seven versions of GT, positivist, postpositivist, constructivist, objectivist, postmodern, situational, and computer assisted GT \cite{Denzin:2010:bookCh21}, with the first three roughly corresponding to the aforementioned classic GT and its two extensions.

In their introductory chapter \cite{Bryant:2010:bookCh1}, Bryant and Charmaz also noted that the original GT did not in anyway reject quantitative methods.
Glaser and Strauss ``intended to show such [qualitative] research projects could produce outcomes of equal significance to those produced by the predominant statistical-quantitative, primarily mass survey methods of the day.'' ``Glaser has always argued that the method applies equally to quantitative inquiry.''
In fact, Glaser and Strauss clearly stated in their 1967 book \cite{Glaser:1967:book}:%
\footnote{The italicized excerpts were highlighted by Glaser and Strauss originally in their book\cite{Glaser:1967:book}.}
``Our position in this book is as follows: there is no fundamental clash between the purposes and capacities of qualitative and quantitative methods or data. ... We believe that \emph{each form of data is useful for both verification and generation of theory,} whatever the primacy of emphasis. ... \emph{In many instances, both forms of data are necessary}.''
Many scholars have adopted mixed-methods approaches featuring both qualitative and quantitative methods.
For example, Brannen included GT as a major method in her 1995 book on mixed methods \cite{Brannen:1995:book}. 
Almost all books thereafter on mixed methods approaches have included GT as component method (e.g., \cite{Creswell:2002:book,Todd:2004:book,Bergin:2018:book}).
In 2010, Johnson et al. proposed a mixed-methods version of GT (MM-GT) for ``connecting theory generation with theory testing'' \cite{Johnson:2010:RiS}.

The contemporary, broadened scope of GT provides an ideal research framework for applying qualitative research methods to the discourse data in the field of VIS, while the GT-based studies can benefit from the research methods familiar to visualization researchers, such as digital data collection, text visualization, and empirical studies.


%% file: sections/4NewWorkflows.tex
\begin{figure*}[t]
    \centering
    \includegraphics[width=135mm]{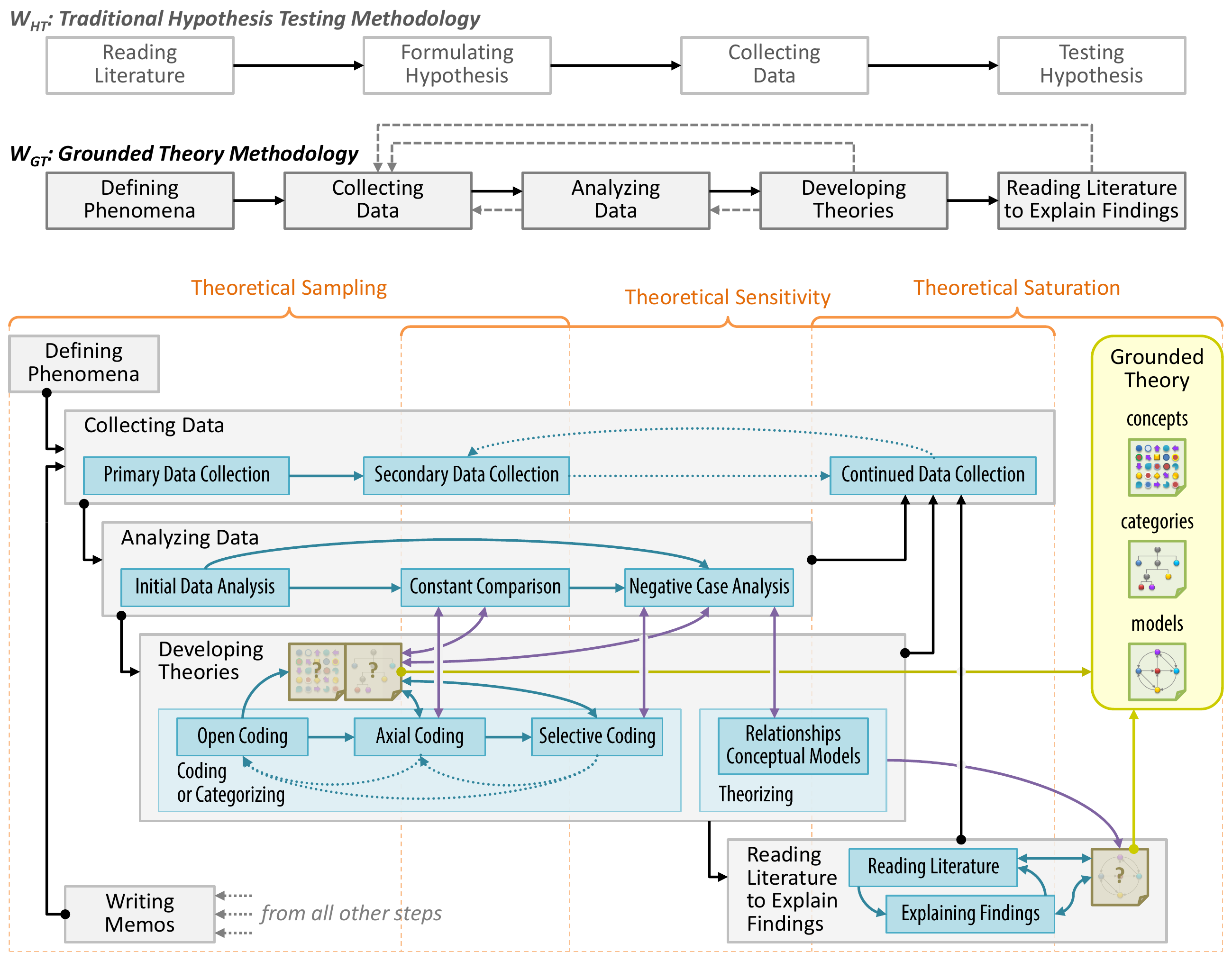}
    \caption{A traditional hypothesis testing workflow is juxtaposed with a GT workflow, which is detailed in the lower half of the figure.}
    \label{fig:GTworkflow}
\vspace{-6mm}
\end{figure*}

\section{Comparing GT Workflows and VA Workflows}
\label{sec:NewWorkflows}
Grounded Theory (GT) advocates research workflows that are different from the traditional hypothesis-testing workflows. As illustrated in Figure \ref{fig:GTworkflow}, the two types of workflows differ in several ways:
\begin{itemize}
    \item A hypothesis-testing workflow $W_{HT}$ begins with reading the literature to formulate a hypothesis, while a GT workflow $W_{GT}$ begins with defining phenomena to be studied without any hypothesis. Note that the notion ``without any hypothesis'' strictly belongs to the Classic GT or Glaserian GT, as the broadened versions of GT, such as Straussian GT and Constructivist GT, considered that it would be natural to have thoughts on the phenomena to be studied.
    \item In $W_{HT}$, one collects data in order to test a hypothesis, while in $W_{GT}$, one collects all collectable data associated with the phenomena. With a known hypothesis, one can normally collect data by focusing on a few independent and dependent variables, where the notion of dependency is usually part of the hypothesis to be tested. Without any hypothesis, one normally has to collect less focused data featuring many variables. 
    \item In $W_{HT}$, collected data represents the sampling of the independent variables that are predefined for a hypothesis, and typically statistical analysis are used to derive measures about predefined dependent variables. The statistical properties associated with both independent and dependent variables are used to reason about the hypothesized relationships among these variables, e.g., scatter plots may be more useful than parallel coordinates plots \cite{Kuang:2012:CGF}, visualization processes benefit from human knowledge \cite{Kijmongkolchai:2017:CGF}.
    
    \hspace{4mm} In $W_{GT}$, as independent and dependent variables are not predefined, it is necessary to analyze the collected data to identify such variables by ``coding'', which attempts to identify from data a number of \emph{concepts} (or \emph{attributes}) and the categories (or types) of each concept (or attribute). Mathematically, a concept (or an attribute) is variable, and a category (or a type) is a valid value of a variable.
    As shown in the lower half of Figure \ref{fig:GTworkflow}, coding and data analysis are two highly integrated processes.
    \item In $W_{HT}$, data is often collected in an organized tabular form, while in $W_{GT}$, data is often collected as free-form textual data or less-organized tabular data (e.g., tabular cells may have inconsistent or missing data values).
    Hence, GT methods are often used to study discourse data.
    \item In Figure \ref{fig:GTworkflow}, $W_{GT}$ appears to be more iterative than $W_{HT}$. Later we will point out that $W_{HT}$ is also iterative but in a different way.
\end{itemize}

\begin{figure*}[t]
    \centering
    \includegraphics[width=135mm]{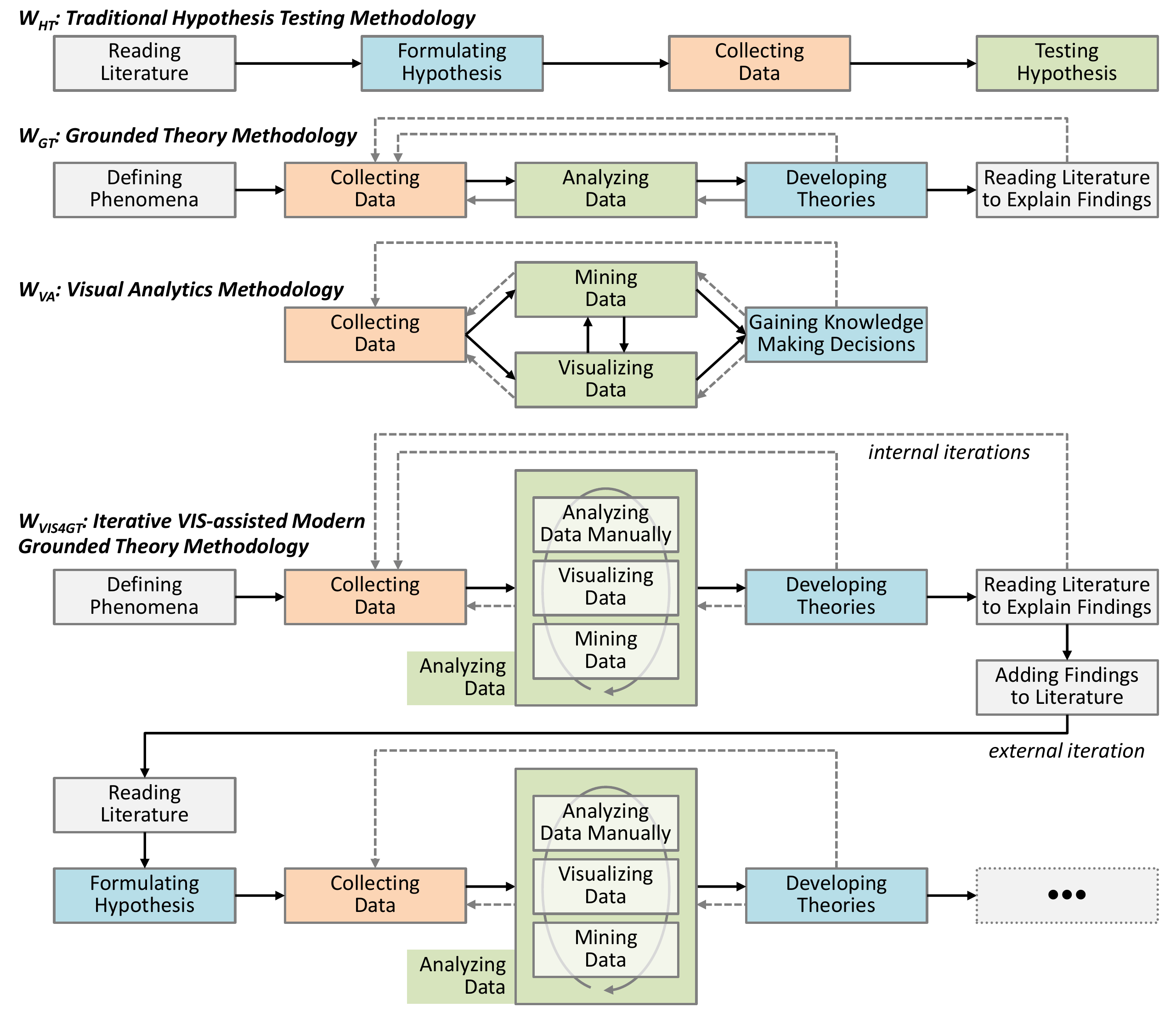}
    \caption{Juxtaposing a traditional hypothesis testing workflow, a GT workflow, and a VA workflow, we can observe that their key processes blocks share similar functionality, e.g., orange blocks for \emph{data collection}, green blocks for \emph{data analysis}, and cyan blocks for data-informed decision making, knowledge acquisition, and hypothesis/theory generation. As the modern GT scope encompasses computer-assisted techniques and hypothesis testing, we can embrace an iterative VIS-assisted GT workflow (VIS4GT) as illustrated in the lower half of this figure.}
    \label{fig:GTVAworkflows}
\vspace{-5mm}
\end{figure*}

Although the terms ``hypothesis'' and ``hypothesis testing'' do not appear in the GT workflow illustrated in Figure \ref{fig:GTworkflow}, it would be overly-strict to assume the absence of hypotheses. For example, to derive a categorization scheme using coding, the researcher has to contemplate different optional schemes and hypothesize their meaningfulness and usefulness. The judgement about theoretical sensitivity and saturation involves hypotheses as to whether new effort for data collection, coding, or data analysis would change the theoretical findings obtained so far.

Although the terms ``statistics'' and ``visualization'' do not appear in the GT workflow illustrated in Figure \ref{fig:GTworkflow}, they are very much part of the data analysis process in GT since it was proposed \cite{Glaser:1967:book}.
As mentioned in Section \ref{sec:NewGT} and Table \ref{tab:GTmethods}, after coding, the researcher typically derives statistical measures for different categories (or types) for each concept (or attribute). Glass and Strauss particularly articulated the need to design appropriate data tables where statistical measures can be aligned intelligently to reveal relations among different categories (or types) and among different concepts (or attributes). Although such visual observation can clearly benefit from visualization techniques, likely and understandably, Glass and Strauss had limited access to visualization techniques at that time (more than five decades ago).

This reminds us of visual analytics (VA) workflows outlined by Keim et al. \cite{Keim:2008:bookCh}. The upper half of Figure \ref{fig:GTVAworkflows} juxtaposes $W_{HT}$ and $W_{GT}$ with a VA workflow $W_{VA}$. We encode process-blocks of a similar function using the same color. In general, ``Testing Hypothesis'' in $W_{HT}$ is a data analysis process, while ``Analyzing Data'' in $W_{HT}$ can be seen as a technically simpler version of ``Mining Data'' and ``Visualizing Data'' in $W_{VA}$.

In the lower half of Figure \ref{fig:GTVAworkflows}, we add the ``Mining Data'' and ``Visualizing Data'' processes in $W_{VA}$ into the ``Analyzing Data'' process in $W_{GT}$. Such an integration is an implementation of Denzin's proposal of computer assisted GT \cite{Denzin:2010:bookCh21}. We refer to such a workflow as VIS-assisted GT methodology, and denote it as $W_{VIS4GT}$. In the right column of Table~\ref{tab:GTmethods}, we elaborate various VA techniques that may be used for assisting different GT methods.

Although the $W_{HT}$ workflows in Figures \ref{fig:GTworkflow} and \ref{fig:GTVAworkflows} do not feature any feedback loop, traditional hypothesis-testing workflows are in fact fairly iterative.
In particular, in psychology and VIS, many hypotheses have been evaluated by multiple empirical studies over a long period, which often feature different experiment designs and different analytical methods, and lead to different conclusions. For example, Kuang et al. conducted a study to test a hypothesis that scatter plots (SPs) are more effective than parallel coordinates plots (PCPs), and obtained a positive conclusion \cite{Kuang:2012:CGF}.
Kanjanabose et al. noticed the ``theoretical sensitivity'' of the conclusion, and designed a study to collect new data for comparing SPs, PCPs and data tables \cite{Kanjanabose:2015:CGF}. The data analysis showed that data tables are more effective than SPs and PCPs for value retrieval tasks, PCPs are more effective for some tasks such as clustering, outlier detection, and change detection. On the topic of perceiving correlation through SPs, there are over 20 hypothesis-testing studies in the literature (see literature review by Sher et al. \cite{Sher:2017:CGF}).

For hypothesis testing, a hypothesis is usually iteratively tested by different research teams in different $W_{HT}$ workflows. We refer to such iterative research workflows as \emph{external} iterations.
For a GT study, it is often the same researcher that conducts multiple iterations as shown in Figure \ref{fig:GTworkflow}. We refer to such iterative research workflows as \emph{internal} iterations.
With the increasing adoption of GT methods in different applications (Figure \ref{fig:GTandVis}), it becomes more common for different research teams to study the same set of phenomena through external iterations.
The lower half of Figure \ref{fig:GTVAworkflows} illustrates both internal and external iterations in such a VIS-assisted GT workflow $W_{VIS4GT}$. Based on the above reasoning, we can conclude:
\begin{enumerate}
    \item A corpus of discourse collected in the field of VIS can be studied using GT in order to infer theories as GT workflows are particularly suitable the free-form textual data in such a corpus;
    \item A proposed visualization theory (e.g., a guideline) should be grounded in the data, and thus GT methods can be used to study such a theory;
    \item When a visualization theory (e.g., a guideline) is considered as a phenomenon to be studied using GT methods, the data collected will likely feature the conditions and contexts in which the theory is applicable or not;
    \item Many VA techniques, including text analysis and visualization, can be used to assist GT research;
    \item Controlled experiments for theory testing are part of GT iteration, if the hypothesis to be tested is a proposed theory originally derived from previous iterations.
\end{enumerate}

As illustrated in Figure \ref{fig:GTandVis}, these conclusions are consistent with the broadened scope of GT.

%% file: sections/5NewWhere.tex
\section{Where are GT Approaches in VIS?}
\label{sec:NewWhere}
From the previous two sections, we can observe that the Classic GT does include statistics and table-based ``visualization'' as part of GT workflows. As depicted in Figure \ref{fig:GTandVis}, the contemporary schools of thought embrace external iterations, computer-assisted GT and empirical studies, all of which hinge on the interpretation that ``open mind'' is ``not empty mind'' \cite{Strauss:1990:book}. Nevertheless, these schools of thought all agree that theories to be postulated and evaluated in GT workflows must be grounded in real-world data.
In other words, there must be a provenance for a sequence of GT workflows, regardless internal and/or external iterations, with which such a theory can be traced back to some real-world data. 
As most GT applications feature qualitative data and free-form textual data, we consider here such discourse, text, and imagery corpora in the following visualization contexts:
\begin{itemize}
    \item \emph{Context A}: Research papers and discourse about them;
    \item \emph{Context B}: Visualization guidelines and discourse about them;
    \item \emph{Context C}: Visualization images and discourse about them;
    \item \emph{Context D}: Discourse about subject developments;
    \item \emph{Context E}: Discourse about schools of thought;
    \item \emph{Context F}: Discourse about design practices;
    \item \emph{Context G}: Discourse about visualization literacy.
\end{itemize}



\subsection{Explicitly-Labelled GT Approaches}
\label{sec:Emplicit}
Visualization researchers have been studying such textual corpora for decades. In some cases, the terms of ``grounded theory'' or ``open coding'' were mentioned in the published work. These explicitly-labelled GT work in VIS include:

\noindent\textbf{Context A.}
Isenberg et al. \cite{isenberg2016visualization} conducted a retrospective analysis of keywords in VIS papers. The use of GT was mentioned.
Kim et al. \cite{kim2021accessible} applied GT to study papers published for the last 20 years on visualization accessibility. Using coding and thematic analysis, they derived a design space for accessible visualization.

\noindent\textbf{Context B.}
Kandogan and Lee \cite{kandogan2016grounded} conducted an extensive GT study on visualization guidelines. They extract some 550 guidelines from books, papers, and blogs, and formulated five high-level concepts (i.e., data, visualization, user, insight, device) as well as qualifiers and sub-concepts. They examined patterns, relations, and co-occurrences among them.
Diehl et al. \cite{Diehl:2020:arXiv} used GT to study the posts on an open discourse forum, VisGuides, specially for discussing visualization guidelines. They followed the contemporary schools of thought in GT, and used text analytics and visualization and empirical studies in their GT workflow. 
Bigelow et al. \cite{bigelow2020guidelines} applied GT to analyze the interviews with data workers and identified concepts and categories, from which they derived guidelines.

\noindent\textbf{Context C.}
Lundgard and Satyanarayan \cite{lundgard2021accessible} used GT to study the semantic content of natural language descriptions that accompany visualization images.
Lee et. al \cite{lee2015people} used GT to analyze different high-dimensional multivariate visualizations that are considered unfamiliar to novice users in order to derive a model of novice users' sense making processes.

\noindent\textbf{Context D.}
Chandrasegaran et al. \cite{Chandrasegaran:2017:CGF} proposed to integrate VA techniques into GT practices in qualitative text analysis.
Sperrle et al. ~\cite{Sperrle2021HCE_STAR} used GT to analyze qualitative feedback and findings on human-related factors that influence human-centered machine learning.

\noindent\textbf{Context E.} \emph{We have not found explicitly-labelled GT work on schools of thought in VIS.}

\noindent\textbf{Context F.}
Rajabiyazdi et al. \cite{rajabiyazdi2020exploring} used GT to study focus group discussions in relation to patient-generated data visualization.
Leon et al. \cite{leon2020mapping} used GT to study survey data on designing global choropleth maps.
The work by Kim et al. \cite{kim2021accessible}, mentioned in the \emph{context A} also falls into this context.

\noindent\textbf{Context G.}
Adar and Lee \cite{adar2020communicative} applied GT to a corpus of communicative visualization and interview data. They built their GT workflows on Bloom's taxonomy for teaching, learning, and assessing \cite{bloom2001taxonomy}. The work demonstrates ``open mind is not empty mind'' -- a contemporary school of thought in GT.

\subsection{Implicitly-labelled or Unlabeled GT Approaches}
\label{sec:Implicit}

Many research papers in VIS involved extensive study of research papers, a collection of which is a text corpus. Hence such work would have featured GT-like workflows, though GT was rarely explicitly mentioned. Consider the aforementioned seven contexts:

\noindent\textbf{Context A.} Many researchers proposed categorization schemes, typologies, taxonomies and ontologies based on extensive literature research. The effort for identifying taxa in a taxonomy or an entity in an ontology is very similar to coding in GT, while determining hierarchical and ontological relations is a form of theorization. Examples of such papers include \cite{borgo2012state,bressa2021s,lam2011empirical,abdul2020survey}.

Many survey papers, which also fall into Context A, produced categorization schemes, taxonomies, ontologies, guidelines, and models that also feature other contexts. For example,

\begin{itemize}
    \item \textbf{Context B.} Borgo et al. \cite{borgo2013glyph} collected a list of guidelines on glyph-based visualization.
    \item \textbf{Context D.} Rogowitz et al., Tovanich et al., Crisan et al. examined the issues of subject development of VIS (e.g., history, diversity, polarization, and biases) through the lens of research papers and conferences \cite{rogowitz2019marshalling,tovanich2021gender,crisan2020passing}.
    \item \textbf{Context E.} A few papers extracted the categories of different schools of thought or arguments in VIS, e.g., \cite{streeb2019visualize,chen2020isms}.
    \item \textbf{Context F.} Many papers proposed design methods, design spaces, and design guidelines based on literature research, e.g., \cite{hall2019design,hullman2018pursuit,cakmak2021multiscale}.
    \item \textbf{Context G.} Liu et al. \cite{liu2021visualization} developed a taxonomy of visualization resources for improving visualization literacy.
\end{itemize}

Some research papers report categorization schemes, typologies, taxonomies, guidelines, methodologies, and conceptual models directly from focus group discussions and interviews, e.g., \cite{meyer2019criteria,parsons2021understanding} in Context F.

%% file: sections/6NewWhyHow.tex
\section{Why and How to Adopt GT Approaches in VIS?}
\label{sec:NewWhyHow}
Visualization and visual analytics (VIS) has always and will continue to provide other disciplines with effective and efficient techniques and tools for supporting data intelligence workflows that humans need to gain information and knowledge from data and to make data-informed decisions.
As a technology, VIS offers a invaluable bridge between machine-centric processes (e.g., digital data collection, statistics, algorithms, computational models, etc.) and human-centric processes (e.g., perception, cognition, knowledge acquisition, decision making, etc.).
As a scientific subject, VIS is an interdisciplinary study of both machine- and human-centric processes in the context of visualization, and in particular, the interactions between these two types of processes.
As Chen and J\"{a}nicke observed through the lens of information theory, when a visualization process is considered as a data communication process, we normally know the encoder well, but not the decoder \cite{Chen:2010:TVCG}. In comparison with the algorithms and software for the ``encoders'', we have rather vague understanding about the ``decoders''. 

Based on above reasoning, the field of VIS should feature a fair amount of social science research. Perhaps it is self-evident that the amount of social science research in VIS has reached the level that is necessary for understanding the ``decoders''. Most social science research in VIS has been conducted in the form of empirical studies, largely related to psychology, one of the many subjects in social science.
As shown in the survey by Abdul-Rahman et al. \cite{Abdul-Rahman:2019:arXiv}, the number of VIS publications on empirical studies has been increasing in recent years.
There is no doubt a need for significantly more social science research in VIS, because:   

\begin{itemize}
    \item \emph{Understanding VIS needs discourse.} ---
    As reviewed in Section \ref{sec:NewRelatedWork}, many discourse papers have played a significant role in VIS, e.g., Van Wijk's values of visualization \cite{van2005value} and Munzner's nested model \cite{munzner2009nested}.
    Many ideas raised in these discourse papers can benefit from community participation in discussions to enrich and improve these ideas over a period. In many cases, new ideas can emerge or be explored before they become mature enough for a publication.
    In VIS, ideally, there are more opportunities for researchers to participate in discussing various discourse topics, through discussion fora as well as formal publications.
    As a scientific discipline, VIS can prosper from discourses, discussions, or debates, which in GT terms, are ``theoretical sampling'', facilitate ``comparative analysis'' and ``negative case analysis'', and can reveal 
    ``theoretical sensitivity'' (Table \ref{tab:GTmethods}).
    \item \emph{Discourse needs abstraction.} ---
    When a scientific or scholarly subject reaches a certain level of maturity, the scientists or scholars will naturally attempt to make abstraction and generalization from empirical evidence and practical experience \cite{chen2020isms}. The researchers in VIS have always pursued and striven for this. GT was developed to assist in abstraction and generalization processes. For example, the most commonly-adopted GT method, coding, is essentially an abstraction method for reducing diverse and complex textual descriptions to a number of categories.
    The intelligent table-based display of statistics proposed by Glaser and Strauss \cite{Glaser:1967:book} was intended to discover abstract and generalizable relationships among different categories. Meanwhile, iterative applications of GT workflows with continuous sampling, comparative analysis, and negative case analysis are quality assurance processes for abstraction and generation.
    \item \emph{Abstraction needs VIS assistance.} ---
    GT workflows have been primarily human-centric processes, and will continue to be in the foreseeable future. Using VIS to assist GT is not in any way intended to replace human-centric processes with machine-centric processes, since visualization and interaction are human-centric processes~\cite{Sperrle2021HCE_STAR}. VIS can significantly improve the efficiency and effectiveness of GT workflows, which will be discussed in detail in the following four subsections.   
    On the one hand, when GT workflows are employed to study VIS discourse, VIS researchers are the domain experts as well as technology-enablers. They will naturally employ different VIS techniques to enrich and improve their GT4VIS workflows. Hopefully, the success of employing VIS techniques can be translated to wide deployment of VIS4GT in many other GT workflows.  
\end{itemize}

The increasing use of VIS techniques will no doubt increase the know-how about GT4VIS and VIS4GT in the field of VIS. In the following four subsections, we describe our experience of GT4VIS and VIS4GT when we used GT methods for analyzing discourse on visualization guidelines. While such experience is not a comprehensive instruction about how to deliver GT4VIS and VIS4GT solutions, we hope that it will prompt VIS colleagues to develop additional and improved GT4VIS and VIS4GT solutions.

\begin{figure*}[t]
    \centering
    \begin{tabular}{@{}c@{\hspace{4mm}}c@{\hspace{4mm}}c@{\hspace{4mm}}c@{}}
        \includegraphics[height=29mm]{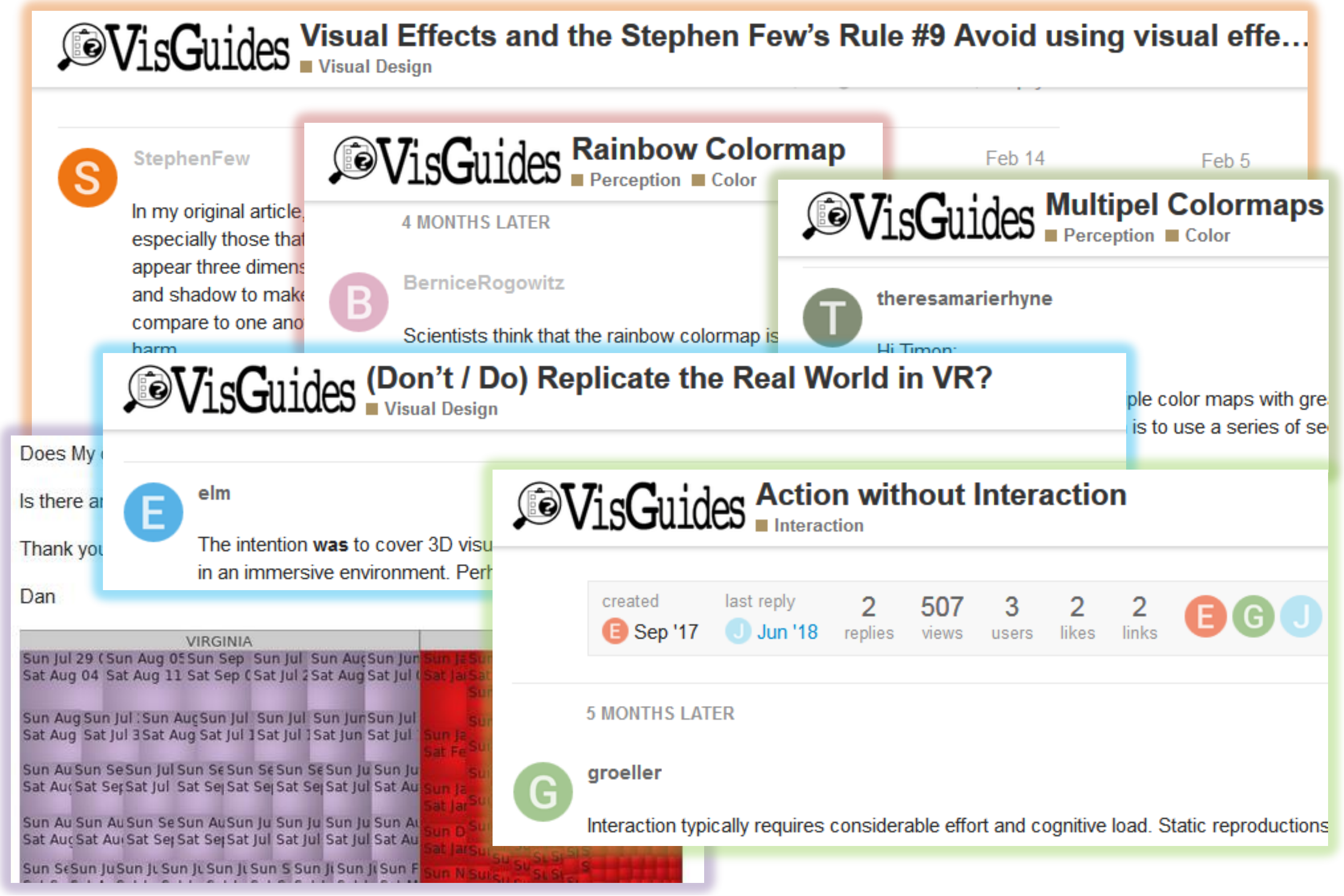} &
        \includegraphics[height=29mm]{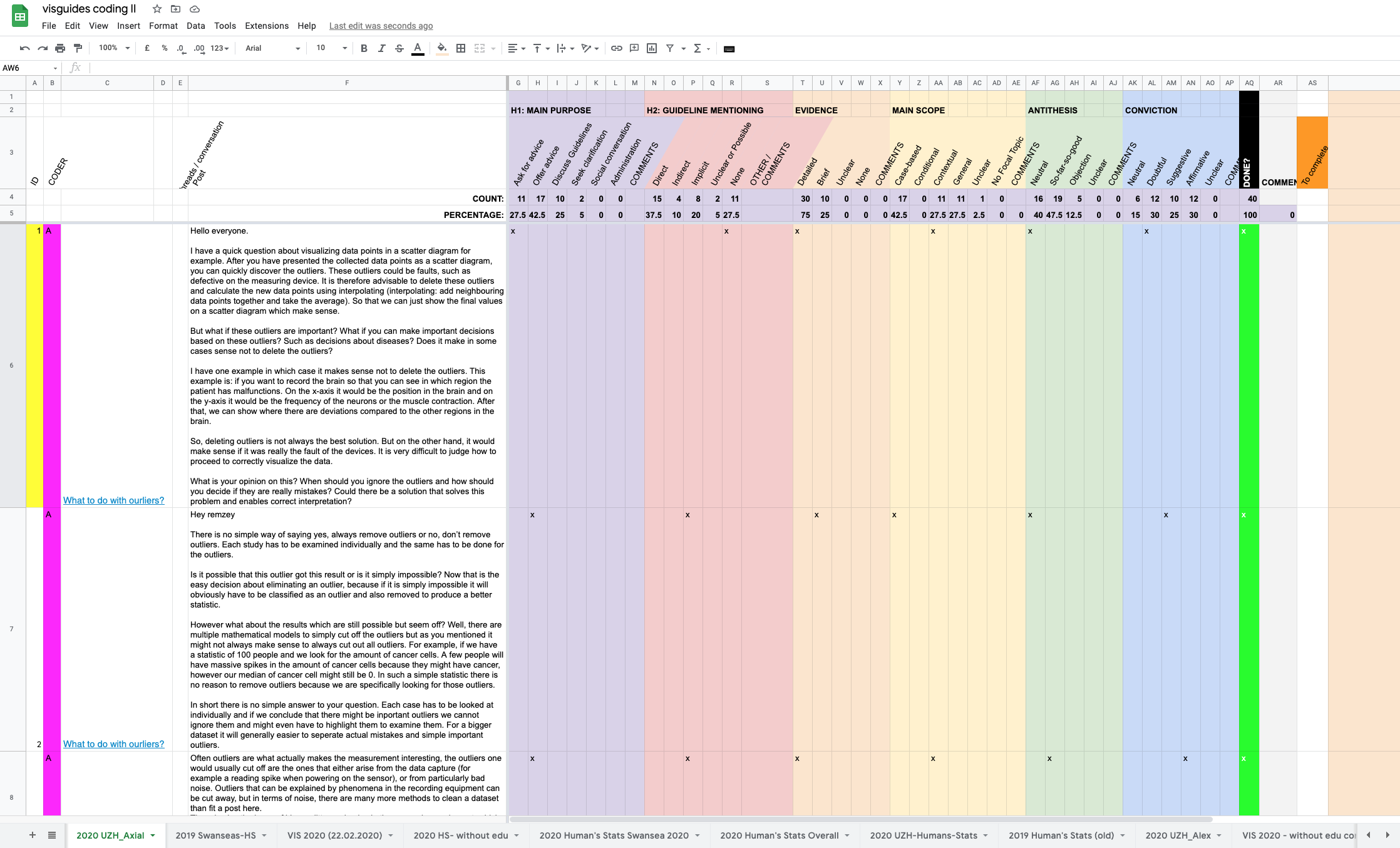} &
        \includegraphics[height=29mm]{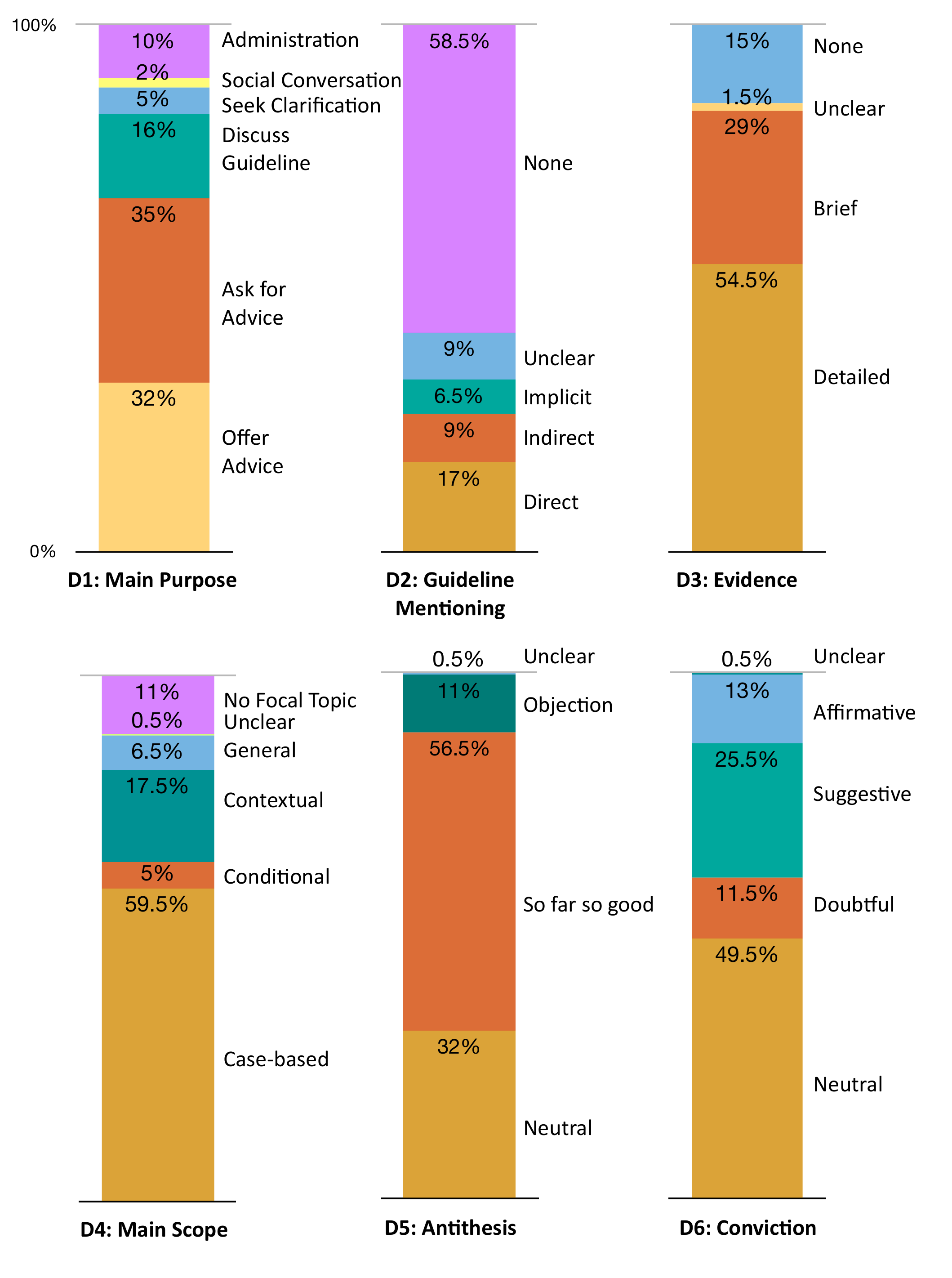} &
        \includegraphics[height=29mm]{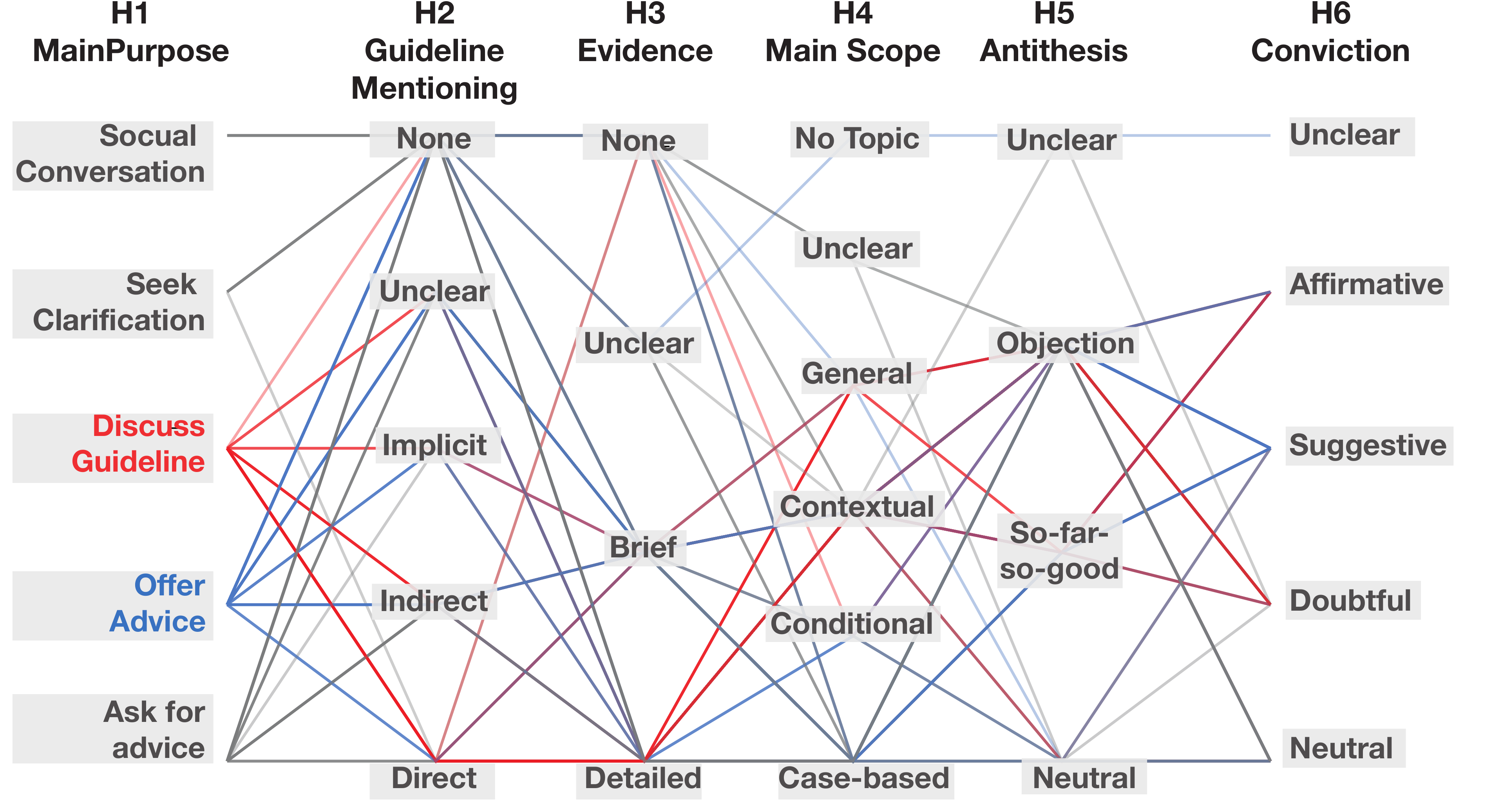}\\
        (a) corpus of posts &
        (b) coding &
        (c) statistics &
        (d) interactive visualization
    \end{tabular}
    \caption{Using statistical graphics and interactive visualization to support coding and categorisation of posts on VisGuides. The parallel coordinates plot in (d) is used for further analysis and theorization. For detailed descriptions about these images, please see \cite{Diehl:2020:arXiv}.}
    \label{fig:Coding}
\vspace{-4mm}
\end{figure*}



\subsection{Open Discourse}
In general, online discourse fora have been more effective in gathering discourse that can be studied using GT methods. VisGuides (\url{https://visguides.org/}) is an online discussion forum dedicated to visualization guidelines~\cite{Diehl:2018:EuroVis}.
It allows registered users to pose questions and offer advice on visualization, propose and recommend guidelines, share positive and negative experience about certain guidelines, report and reason about successes, failures, and conflicts of guidelines, and suggest ways to refine guidelines.

A guideline essentially defines a causal relation that an action $A$ will cause a consequence $B$ under condition $C$. In many cases, guidelines are expressed as ``one should do $A$'', implying a negative consequence. In most cases, the condition $C$ is yet to be precisely defined. Such a guideline would be considered as a postulated theory in social science and many other fields. With hundreds of visualization guidelines available in the literature \cite{kandogan2016grounded}, it would take a very long time for all proposed guidelines to be investigated thoroughly. 

VisGuides serves as a digital platform to supply both the \textit{postulated theories}, i.e., the visualization guidelines to be developed and investigated, and the \textit{data}, i.e., the discourses that can be used to generate, evaluate, and refine the theories. 

From the GT perspective, VisGuides provides a mechanism for community-based ``theoretical sampling'', facilitating the evaluation of ``theoretical sensitivity''. For example, when the failure of a guideline under a specific condition is reported, it is a ``negative case'' to be analysed, indicating that the guideline is still theoretically sensitive.

VisGuides was launched on September 27, 2017, and according to the latest count, the forum received 577 posts on December 1, 2021, which include contributions from users with different visualization backgrounds, ranging from undergraduate students to highly-respected visualization experts (e.g., in alphabetic order, Niklas Elmqvist, Stephen Few, Eduard Gr\"{o}ller, Robert Kosara, Liz Marai, Theresa-Marie Rhyne, and Bernice Rogowitz, among others).

VisGuides has been also used for visualization education, providing a platform for students to discuss their visual designs, seek advice on visualization guidelines, and offer their discourse on individual guidelines \cite{DFT:21}.  

\subsection{Support Coding}
As mentioned in Section \ref{sec:NewWhere}, coding has been used by many VIS researchers in categorizing discourse data and other complex data. In the case of discourse data from VisGuides, we have applied open coding, axial coding, and selective coding to the data. Figure \ref{fig:Coding} illustrates the main steps for transforming raw discourse data (a) to coded concepts and categories in spreadsheets (b), which is then transformed to statistical graphics (c).
The detailed descriptions of these steps can be found in an arXiv report by Diehl et al. \cite{Diehl:2020:arXiv}.

As mentioned in earlier sections, Glaser and Strauss considered statistics as a post-coding step and used data tables to display statistical measures \cite{Glaser:1967:book}. Today, most users can easily create statistical graphics for such data tables. While tables can display data more precisely, statistical graphics can enable users to overview and compare many statistical measures quickly.
For example, during coding, GT researchers frequently encounter a dilemma whether to merge two or a few proposed categories. The relative sizes of different categories (i.e., the numbers of data samples in different categories) are usually important factors that influence the decision. Statistical graphics can enable data comparison more efficiently than data tables.    

\subsection{Enable Analysis and Theorization}
For reasoning about the relationships among coded categories, the GT proposers, Glaser and Strauss, recommend aligning the numbers in the data table intelligently. This is indeed a multivariate data visualization problem. For example, Figure \ref{fig:Coding}(d) shows a parallel coordinates plot for investigating the relationships among different categories under concepts (i.e., variables) H1, H2, ..., H6. The thickness or opacity and the lines encodes the percentage/proportions of a given category and dimension. A GT researcher can interactively brush categories (e.g., the categories of ``Discuss Guideline'' (red) and ``Offer Advice'' (blue), and observe the strength of their relationships with other categories. There is little doubt that interactive visualization can significantly improve the process of aligning table items intelligently.

There are many other multivariate analytical and visualization techniques that can be used for supporting GT, especially, the processes of comparative analysis, negative case analysis, and theorization. For example, one can have tree visualization for proposed taxonomy; graph visualization for proposed ontology, association, and causality; VA-based clustering and anomaly detection for comparative analysis and negative case analysis; and coordinated multiple views for complex analysis and theorization in GT workflows.

\begin{figure*}[t]
  \centering
  \begin{tabular}{@{}c@{\hspace{4mm}}c@{\hspace{4mm}}c@{}}
    \includegraphics[height=37mm]{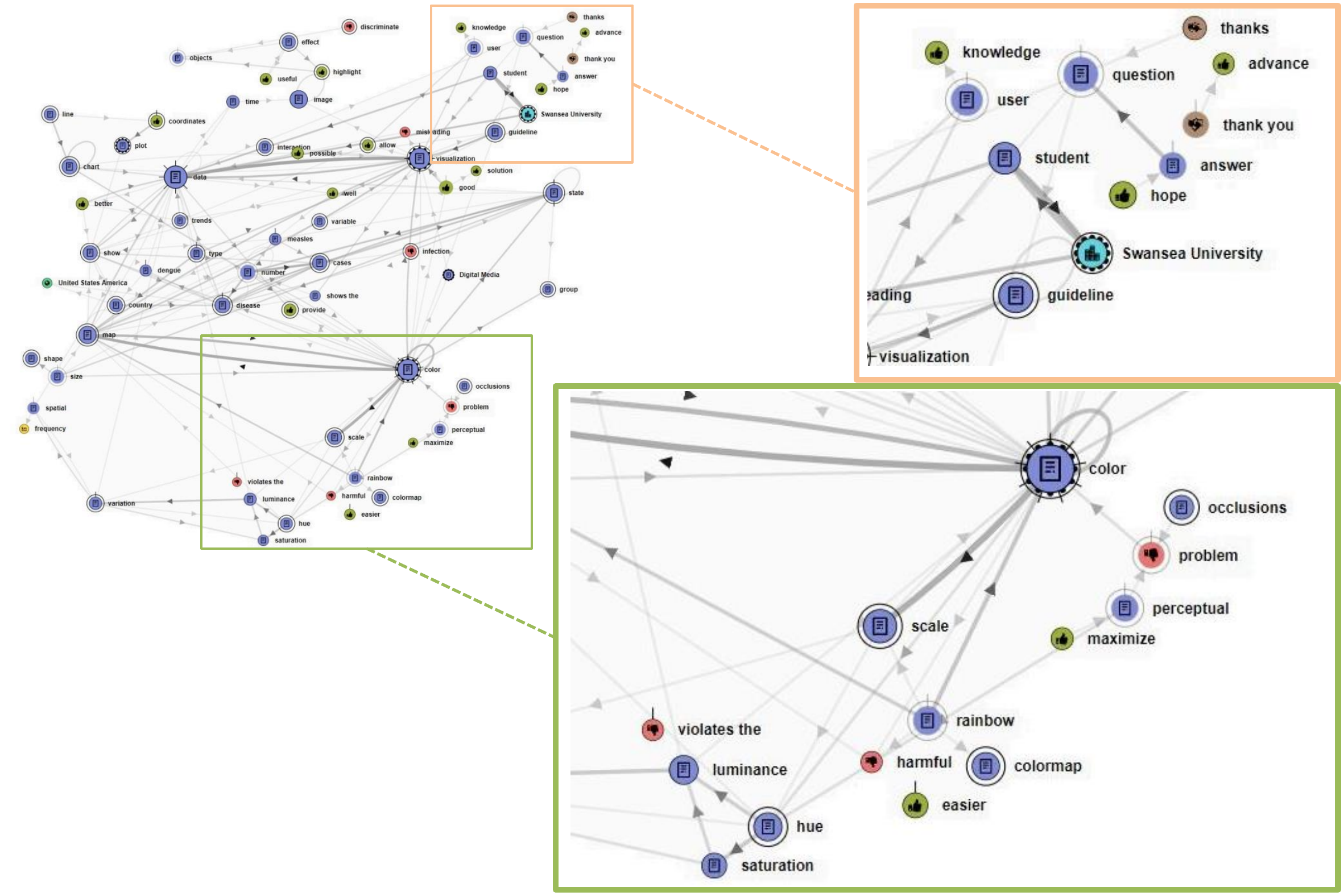} &
    \includegraphics[height=37mm]{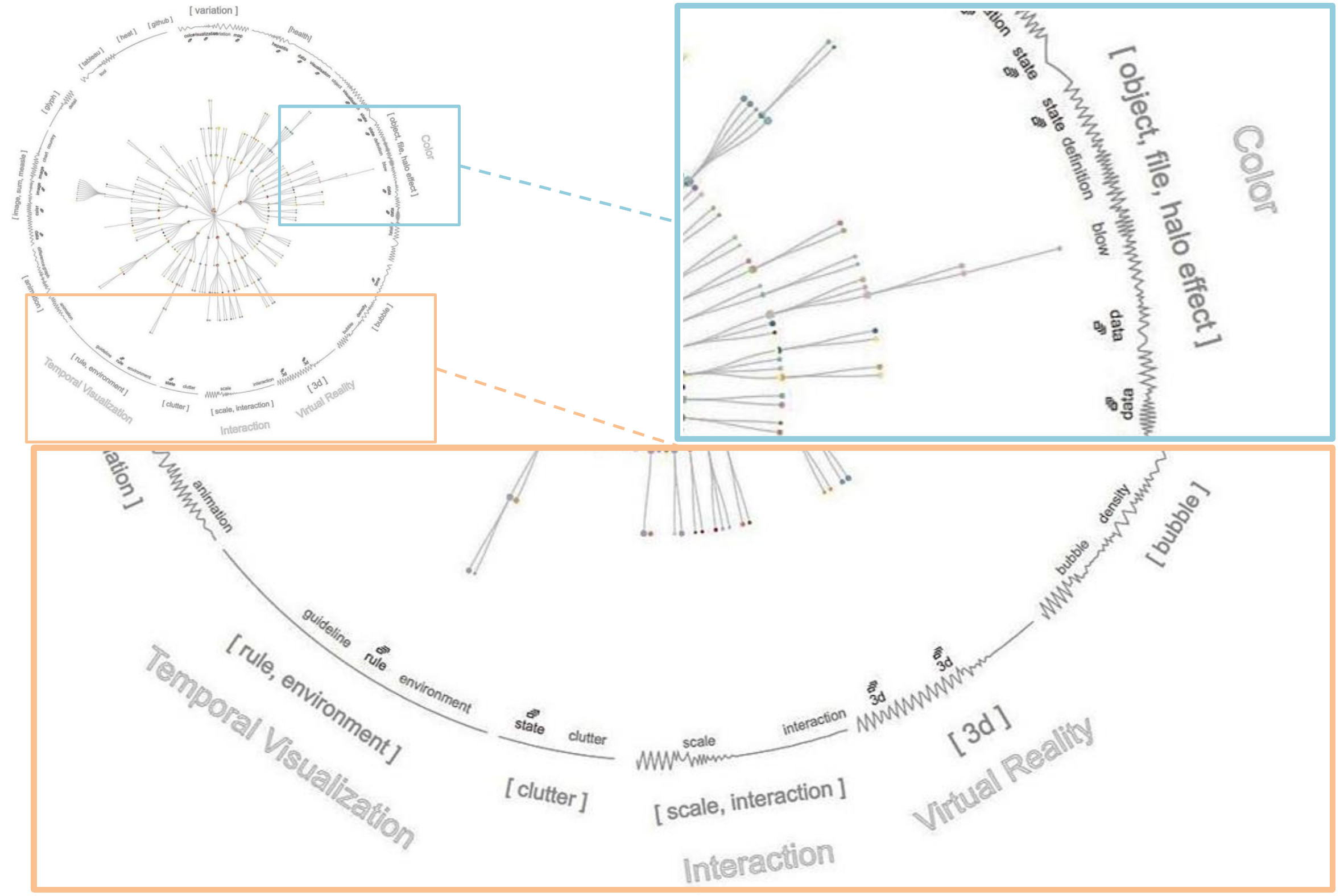} &
    \includegraphics[height=37mm]{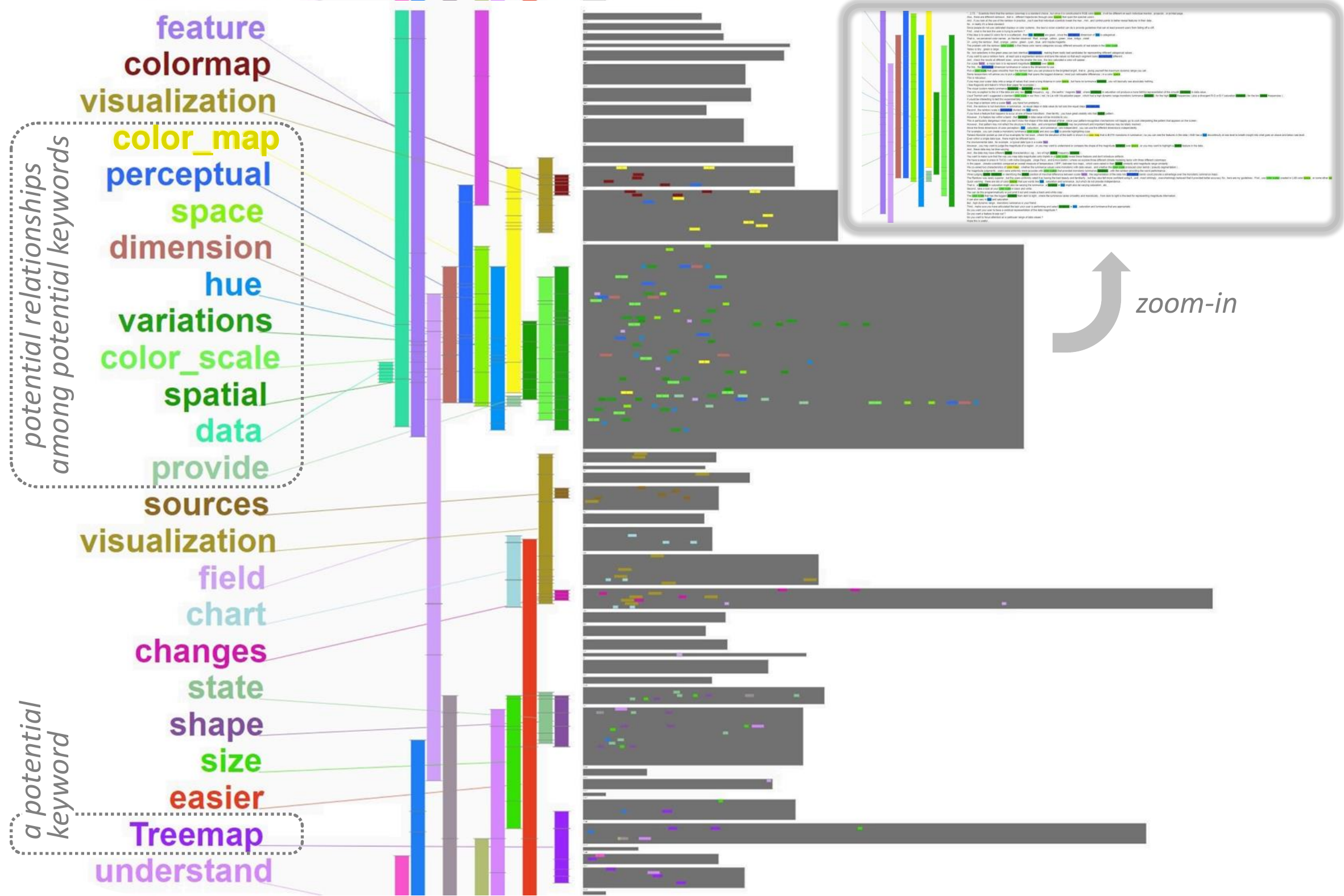}\\
    \small{(a) a named entity graph~\cite{El-Assady2017}} &
    \small{(b) a topic tree~\cite{el2018visual}} &
    \small{(c) a Lexical Episode Plot~\cite{Gold15}}
    \end{tabular}
    \caption{Three different visual text analytics techniques were used in supporting theoretical sampling during the workflow for studying the discourse on VisGuides. For a high resolution version of the images, see \cite{Diehl:2020:arXiv}; and for more detailed explanations, see \cite{2019-lingvis}.}
    \label{fig:TextVis}
    \vspace{-2mm}
\end{figure*}




\subsection{Facilitate Theoretical Sampling}
Because the GT processes in earlier parts of a GT workflows typically focus on coding free-form textual data for which categorization schemes are not yet defined, close reading of raw data has always been a dominant mechanism. We do not suggest replacing automated text analysis with close reading in GT workflows.

However, all those who have experienced a coding exercise appreciate the time-consuming nature of close reading. Although the coding activities are required to be ``open-minded'', in practice, contemplating and exploring many optional categorization schemes is usually hindered by lack of human and time resources. The difficulties in contemplating and exploring options fundamentally hamper the GT principle of ``Theoretical Sampling'', i.e., one cannot afford to sample many options.

Text analysis and visualization is a family of VA techniques that can significantly alleviate the difficulties in contemplating and exploring optional categorization schemes.
With such techniques, a GT researcher can contemplate a categorization scheme using a collection of keywords that are mapped to different concepts and categories.
The text algorithms can scan the raw text and attempt to determine the category (or categories) of the text concerned.
Text visualization can quickly display the results, enabling the GT researcher to judge if the proposed categorization scheme is promising or not, and if further close reading should be performed.
In this way, the GT researchers can explore more optional categorization schemes, as required by the GT principle of ``Theoretical Sampling''.

In addition to statistical graphics for displaying the measures of the proposed categories, graph visualization can be used to depict the relationships between the categories.
For example, Figures \ref{fig:TextVis}(a,b) show two visual representations used in conjunction with category analysis (i.e., referred to as entity analysis in text mining).
 
Text visualization can also support close reading. Figure \ref{fig:TextVis}(c) shows a lexical episode plot ~\cite{Gold15} used to depict the evolution of the discourse over time. The analytical algorithm automatically detects compact chains of $n$-grams and highlights them beside an abstract overview of the complete text. We can interactively zoom-in to the interesting text areas for close reading.
Large versions of Figure \ref{fig:TextVis}, plus further details about using text VA to analyze VisGuides discourse, can be found in an arXiv report by Diehl et al. \cite{Diehl:2020:arXiv}.

\begin{figure*}[t]
  \centering
  \begin{tabular}{@{}c@{}c@{\hspace{4mm}}c@{}c@{}}
    \includegraphics[height=24mm]{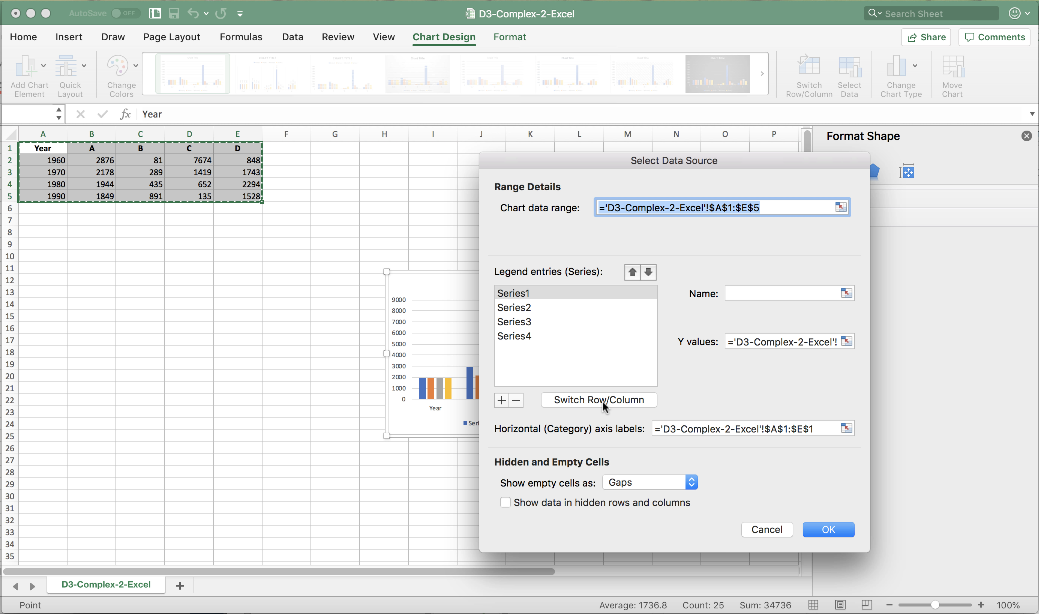} &
    \includegraphics[height=24mm]{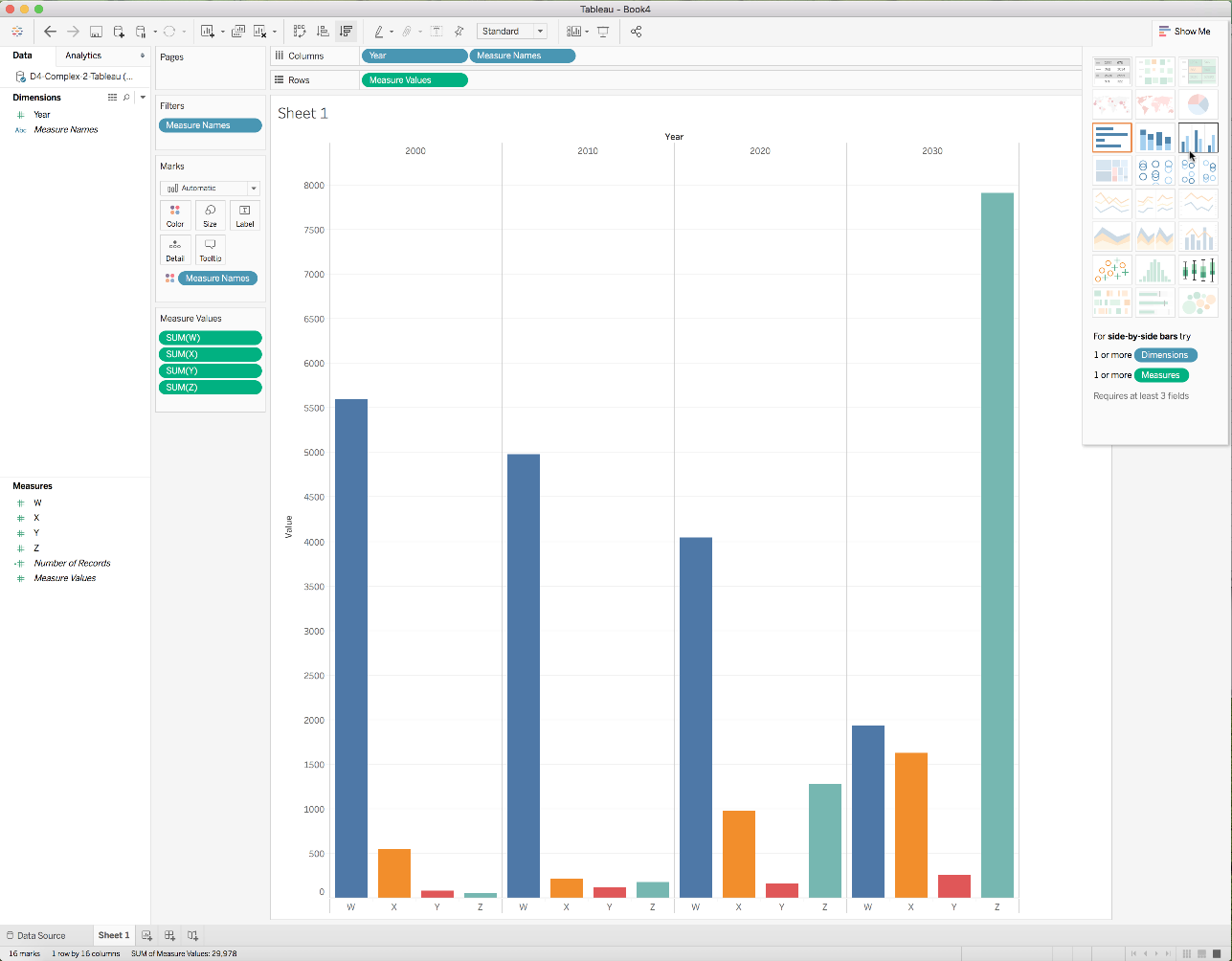} &
    \includegraphics[height=24mm]{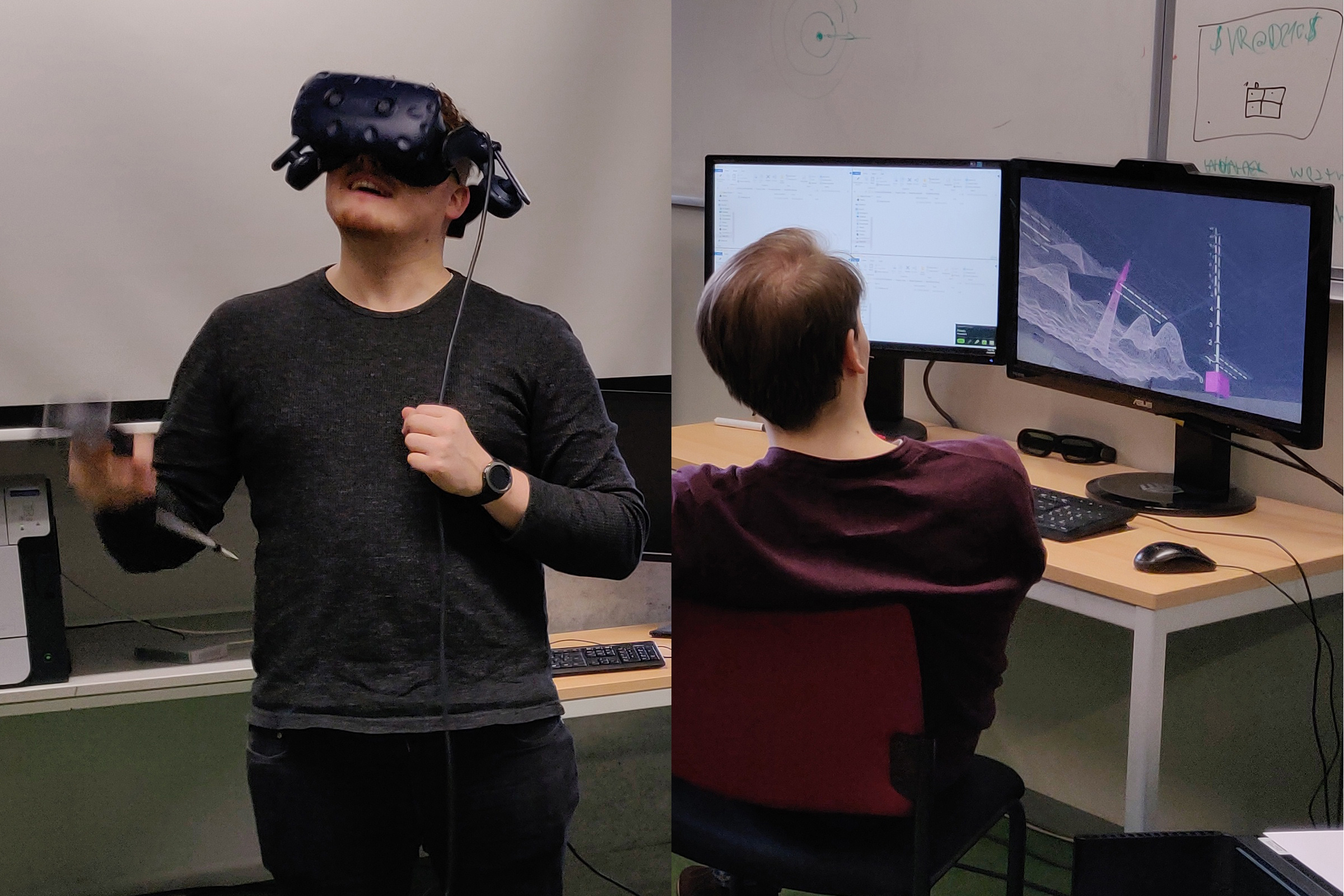} &
    \includegraphics[height=24mm]{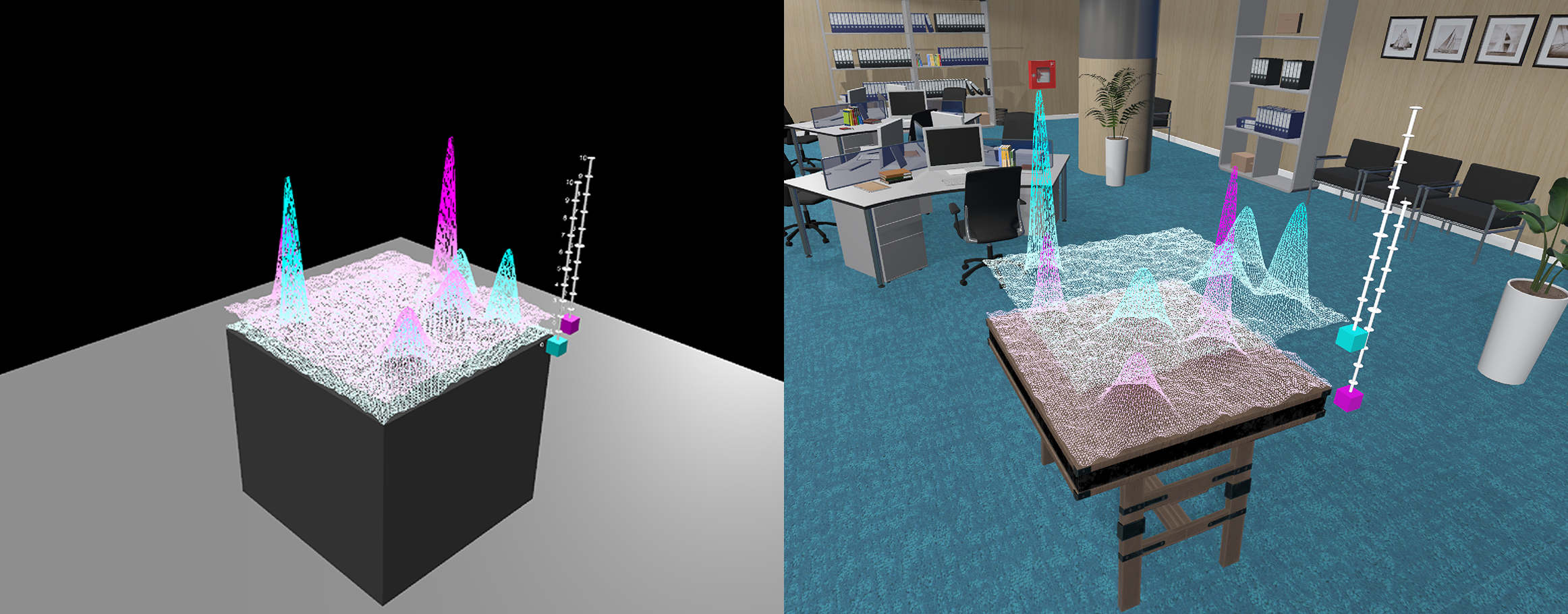}\\
    \multicolumn{2}{c}{(a) two stimuli (MS-Excel and Tableau)} &
    \multicolumn{2}{c}{(b) the VR experimental setting and two stimuli (plain and office scene)} \\
  \end{tabular}
  \caption{The data collection settings for the two empirical studies (Section \ref{sec:ThSensitivity}), which were designed to address the theoretical sensitivity of two visualization guidelines: (a) ``Use interaction in visualization sparsely and cautiously'', and (b) ``Don't replicate the real world in VR''.}
 \label{fig:VR}
\vspace{-5mm}
\end{figure*}

\subsection{Address Theoretical Sensitivity}
\label{sec:ThSensitivity}
Theoretical Sensitivity is another important principle of GT. It obligates GT researchers to address uncertainty in a postulated theory generated in the previous iterations by continuing the iterative workflow, including collecting more data if necessary. One contention in the early schools of thought of GT was whether empirical studies would be considered as GT methods. On the one hand, empirical studies are typically designed to test certain hypotheses, which appeared not to be open-minded. On the other hand, apart from the first iteration, all other iterations in a GT workflow must be influenced by the findings in the previous iterations. Such findings are hypotheses until theoretical saturation.
Many researchers argued for the extension of the traditional GT to include controlled empirical studies in order to evaluate and validate hypotheses \cite{Henwood:1992:BJP,Henwood:2003:BookCh,Badreddin:2012:USER}.
As mentioned in Section \ref{sec:NewGT}, Strauss and Corbin reinterpreted ``open mind'' as not ``empty mind'' \cite{Strauss:1998:book}, opening the door for the inclusion of empirical studies in GT as mixed method GT.

During the close reading of VisGuides discourse, we noticed two threads of posts suggesting some VIS guidelines are still theoretically sensitive (i.e., not yet conclusive). One thread \cite{east101:2017:VisGuides} was initiated by VisGuides user \texttt{east101} to discuss a guideline \textit{``Use interaction in visualization sparsely and cautiously ...''}~\cite{Groller:2008:Dagstuhl}. In the forum, user \texttt{groeller} explained that \textit{``interaction typically requires considerable effort and cognitive load.''}. User \texttt{jamesscottbrown} later presented a contrasting argument: \textit{``... there is generally little cost (to the user) associated with adding such interactivity to a visualisation ...'}'. In order to address the theoretical sensitivity shown in this thread of discussion, we conducted an empirical study to collect more targeted data. 
Another thread was initiated by VisGuides user \texttt{matthias.kraus} on the topic ``\textit{(Don't / Do) Replicate the Real World in VR?}''~\cite{kraus:2018:VisGuides}.
In this thread, the user asked whether the guideline \textit{``Rule \#7: Don't Replicate the Real World''}~\cite{Elmqvist:2017:blog} would still hold in a reality (VR) context.
The thread received a fair amount of discussion, among others, a comment by Nilkas Elmqvist (user \texttt{elm}), confirming that his guideline was meant to include VR.
Similarly we noticed the theoretical sensitivity, and conducted an empirical study to collect more targeted data. Figure \ref{fig:VR} shows some stimuli used in the study.

Both empirical studies are reported in the arXiv report by Diehl et al. \cite{Diehl:2020:arXiv}. In the context of this paper, these two studies show that the existing knowledge and experience of conducting empirical studies in VIS can enrich GT workflows for studying VIS phenomena. In these two cases, the phenomena are about two visualization guidelines, and the discourse data collected initially through VisGuides exhibits a high-level of theoretical sensitivity. Further data collection and analysis is necessary, and empirical studies enable us to collect data on a narrowly-focused topic more quickly than a discourse forum.

%% file: sections/7NewConclusions.tex
\section{Discussions and Conclusions}
\label{sec:NewConclusions}
In this paper, we present our qualitative analysis of the role of visualization and visual analytics (VIS) in grounded theory (GT) workflows. Our study confirms the following findings:
\begin{enumerate}
    \item GT methods are particularly useful for studying discourse data. VIS is an interdisciplinary field involving the study of humans, and can benefit from adopting social science research methods.
    \item Discourse research is not uncommon in VIS and has significant impact on VIS development. More discourse, including open discourse fora, should be encouraged and facilitated.
    \item Since the birth of GT, statistics have been part of GT workflows. Contemporary schools of thought in GT embrace computer-assisted GT.
    GT workflows will welcome visual analytics.
    \item The early idea of arranging data tables intelligently was an attempt of analyzing multivariate data using a crude form of visualization. Interactive VIS can significantly enhance such analytical processes.
    \item The requirement for iterative theoretical sampling in GT logically implies hypothesis-based iterations, since it is not feasible to ensure totally open-mindedness in succeeding iterations. Contemporary schools of thought in GT embrace empirical studies as part of mixed method GT (MMGT).
\end{enumerate}

\noindent\textbf{Our recommendation.} The field of VIS is naturally suitable for developing research agenda of GT4VIS as well as VIS4GT. We can start by incorporating more VIS techniques and empirical studies in GT4VIS workflows, and aspire to provide a family of VIS4GT techniques to the broad GT communities in other disciplines.